\documentclass[pre,twocolumn,showpacs,superscriptaddress]{revtex4}
\usepackage{graphicx,color,amsmath,amsthm,amssymb,bm,array}
\graphicspath{{./Figs/}}
\usepackage{dsfont}

\begin{document}
 
\title{Ring bursting behavior en route to 
turbulence in quasi two-dimensional Taylor-Couette flows}

\author{Sebastian Altmeyer} \email{sebastian.altmeyer@ist.ac.at}
\affiliation{Institute of Science and Technology Austria,
3400 Klosterneuburg, Austria}

\author{Younghae Do} \email{yhdo@knu.ac.kr}
\affiliation{Department of Mathematics,
KNU-Center for Nonlinear Dynamics, 
Kyungpook National University, Daegu, 702-701, Korea}  
 
\author{Ying-Cheng Lai} 
\affiliation{School of Electrical, Computer and Energy Engineering,
Arizona State University, Tempe, Arizona 85287, USA} 
 


\date{\today}

\begin{abstract}

We investigate the quasi two-dimensional Taylor-Couette system in the  
regime where the radius ratio is close to unity - a transitional 
regime between three and two dimensions. By systematically 
increasing the Reynolds number we observe a number of standard
transitions, such as one from the classical Taylor vortex flow (TVF) 
to wavy vortex flow (WVF), as well as the transition to 
fully developed turbulence. Prior to the onset of turbulence 
we observe intermittent burst patterns of localized turbulent 
patches, confirming the experimentally observed pattern of
very short wavelength bursts (VSWBs). A striking finding
is that, for Reynolds number larger than the onset of VSWBs, a 
new type of intermittently bursting behaviors emerge: burst
patterns of azimuthally closed rings of various orders. We call
them {\em ring-burst} patterns, which surround the cylinder 
completely but remain localized and separated by non-turbulent 
mostly wavy structures in the axial direction. We use a number of quantitative
measures, including the cross-flow energy, to characterize the
ring-burst patterns and to distinguish them from the background 
flow. The ring-burst patterns are interesting because it does not
occur in either three- or two-dimensional Taylor-Couette flow: it 
occurs only in the transition, quasi two-dimensional regime of the
system, a regime that is less studied but certainly deserves further 
attention so as to obtain deeper insights into turbulence.

\end{abstract}
\maketitle

\section{Introduction} \label{sec:intro}

Characteristics of turbulence in three- and two-dimensional
flows are typically quite distinct. For example, in three-dimension flows,
the energy spectrum of fully-developed turbulence obeys the well-known
Kolmogorov 1941 scaling law~\cite{Kolmogorov:1941} of $k^{-5/3}$, while 
in two dimensions the scaling~\cite{KM:1980,Lesieur:book} is $k^{-3}$.  
Although two-dimensional flow systems offer great advantages from the 
standpoint of computation and mathematical analysis as compared with
three-dimensional flows, the former are not merely a kind of toy model 
of turbulence. In fact, two-dimensional turbulence is relevant to the 
dynamics of oceanic currents, origin of the ozone hole through mixing 
of chemical species in the polar stratosphere, the existence of polar 
vortex, strong eddy motions such as tropical cyclones, and other
large-scale motions of planetary atmospheres~\cite{DL:1993,Waughetal:1994}.

Turbulence is arguably one of the most difficult problems in science
and engineering, and the vast literature on turbulence 
was mostly focused on three or two dimensions~\cite{Frisch:book}. To gain
new insights into turbulence, it is of interest to study the {\em transitional
regime} between three and two dimensions. In such a regime, properties of
both three- and two-dimensional flows are relevant, and one naturally 
wonders whether there are any unexpected features associated with, for instance,
transition to turbulence~\cite{GS:1975,GB:1980}. The purpose of this paper 
is to report a new phenomenon in a paradigmatic quasi-two-dimensional
system: the Taylor-Couette flow~\cite{Taylor:1923} with gap 
between the inner and outer cylinders so narrow that the system is neither
completely three-dimensional nor exactly two-dimensional. In fact, this 
regime has not been investigated systematically previously. 
The new type of intermittent dynamics occurs {\em en route} to turbulence 
as the Reynolds number is increased.

The Taylor-Couette system, a flow between two concentric rotating cylinders,
has been a paradigm in the study of complex dynamical behaviors of fluid flows, 
especially turbulence~\cite{Coles:1965,Sny69b,DS:book,ALS86,CSA07,Tagg:1994,
CI:book,CoMa1996,AH:2010}. The flow system can exhibit a large variety 
number of ordered and disordered behaviors in different parameter 
regimes. For cylinders of reasonable length, the effective dimensionality 
of the system is determined by a single parameter - the ratio between the 
radii of the two cylinders. If the ratio is markedly less than unity, the
flow is three dimensional. As the ratio approaches unity, the flow becomes
two dimensional. Most previous studies focused on the setting where the radius 
ratio is below, say about 0.9, the so-called wide-gap regime~\cite{ALS86},
or when the ratio approaches unity so that the geometry is locally
planar, resulting in an effectively two-dimensional Couette flow~\cite{FaEc2000}. 
Here, we are interested in the narrow-gap case, where the radius ratio 
is close to unity but still deviates from it so that the flow is 
quasi two-dimensional. To be concrete, we fix the radius
ratio to be $0.99$ and, without loss of generality, restrict our study 
to the case where the outer cylinder is stationary. In fact, regardless 
of whether the outer cylinder is rotational or stationary, transition to
turbulence can occur with increasing Reynolds number. In particular, 
for systems of counter-rotating cylinders, an early work~\cite{Coles:1965} 
showed that transition to turbulence can be sudden as the Reynolds number 
is increased through a critical point. For fixed outer cylinder, the 
transition from laminar flow to turbulence can occur through a sequence
of instabilities of distinct nature~\cite{FSG79,CSA07}.
 
For Taylor-Couette system of counter-rotating cylinders, spatially 
isolated flow patterns, the so-called localized patches, 
can emerge through the whole fluid domain and decay~\cite{ALS86,CA96}.
Depending on the parameters, the localized patches can be laminar or
more complex patterns such as inter-penetrating spirals. In the wide-gap
regime (three-dimensional flow), numerical simulations~\cite{Don2007} 
revealed the existence of so-called G\"ortler vortices~\cite{Goe1954},
small scale azimuthal vortices that can cause streaky structures and form 
herringbone-like patterns near the wall. Localized turbulent behaviors
can arise when the G\"ortler vortices concentrate and grow at the outflow 
boundaries of the Taylor vortex cell~\cite{Don2007}. 

For narrow gap (quasi two-dimensional) flows, there was experimental 
evidence of the phenomenon of very short wavelength bursts 
(VSWBs)~\cite{CSA07}. One contribution of our work is clear computational 
demonstration of VSWBs. Remarkably, we uncover a class of solutions in
quasi two-dimensional Taylor-Couette flow {\em en route} to turbulence. 
These are localized, irregular, intermittently bursting, azimuthally closed
patterns that manifest themselves as various rings located along the axial 
direction. For convenience, we refer to the states as ``{\em ring-bursts}.'' 
The number of distinct rings can vary depending on the parameter 
setting but their extents in the axial direction are similar. The ring 
bursts can occur on some background flow that is not necessarily 
regular. For example, in typical settings the background can be wavy 
vortex flows (WVFs) with relatively high azimuthal wave numbers. Because
of the coexistence of complex flow patterns, to single out ring bursts
is challenging, a task that we accomplish by developing an effective 
mode separation method based on the cross-flow energy. We also find that
ring bursts are precursors to turbulence, {\em signifying a new route to
turbulence} uniquely for quasi two-dimensional Taylor-Couette flows. 
To our knowledge, there was no prior report of ring bursts patterns or 
likes. This is mainly due to the fact that the quasi two-dimensional regime 
is a less explored territory in the giant landscape of turbulence research. 
It would be interesting to identify precursors to turbulence
in quasi two-dimensional flow systems in general.     

In Sec.~\ref{sec:methods}, we outline our numerical method and describe a 
number of regular states in quasi two-dimensional Taylor-Couette flows. In 
Sec.~\ref{sec:RB_characterization}, we present our main results: numerical 
confirmation of experimentally observed VSWBs and more importantly, 
identification and quantitative confirmation of intermittent ring bursts 
as precursors to turbulence. In Sec.~\ref{sec:conclusions}, we present
conclusions and discussions.

\section{Numerical method and basic dynamical states of quasi 
two-dimensional Taylor-Couette flow} \label{sec:methods}

\subsection{Numerical method} \label{subsec:method}

The Taylor-Couette system consists of two independently rotating 
cylinders of finite length $L$ and a fluid confined in the annular 
gap between the two cylinders. We consider the setting in which the
inner cylinder of radius $R_i$ rotates at angular speed $\Omega$ and 
the outer cylinder of radius $R_o$ is stationary. The end-walls 
enclosing the annulus in the axial direction are stationary and the 
fluid in the annulus is assumed to be Newtonian, isothermal and 
incompressible with kinematic viscosity $\nu$. Using the gap width 
$d=R_o-R_i$ as the length scale and the radial diffusion time (DT) 
$d^2/\nu$ as the time scale, the non-dimensionalized Navier-Stokes 
and continuity equations are
\begin{equation} \label{EQ:NSE}
\partial_t \bm{u}+(\bm{u}\cdot \nabla)\bm{u}= -\nabla p + \nabla^2
\bm{u}, \qquad \nabla \cdot \bm{u}=0,
\end{equation}
where $\bm{u}=(u_r,u_{\theta},u_z)$ is the flow velocity field in cylindrical 
coordinates $(r,\theta,z)$, the corresponding vorticity is given
by $\nabla \times \bm{u} = (\xi,\eta,\zeta)$, and $r$ is the normalized 
radius for fluid in the gap ($0 \le r\le 1$). The three relevant 
parameters are the Reynolds number $Re=\Omega_i R_i d/\nu$, the 
radius ratio $R_i/R_o=0.99$, and the aspect ratio $\Gamma \equiv L/d=44$.
The boundary conditions on the cylindrical surfaces are of the non-slip
type, with $\bm{u}(r_i,\theta,z,t)=(0,Re,0)$, $\bm{u}(r_o,\theta,z,t)=(0,0,0)$, 
where the non-dimensionalized inner and outer radii are $r_i= R_i/d$ and 
$r_o= R_o/d$, respectively. The boundary conditions in the axial direction
are $\bm{u}(r,\theta,\pm 0.5 \Gamma,t)=(0,0,0)$.

We solve Eq.~\eqref{EQ:NSE} by using the standard second-order time-splitting 
method with consistent boundary conditions for the pressure~\cite{HuRa98}. 
Spatial discretization is done via a Galerkin-Fourier expansion in $\theta$ 
and Chebyshev collocation in $r$ and $z$. The idealized boundary conditions 
are discontinuous at the junctions where the stationary end-walls meet the 
rotating inner cylinder. In experiments there are small but finite gaps at 
these junctions where the azimuthal velocity is adjusted to zero. To achieve
accuracy associated with the spectral method, a regularization of the 
discontinuous boundary conditions is implemented, which is of the form
\begin{equation}
u_{\theta}(r,\theta,\pm0.5\Gamma,t)= Re\lbrace\exp([r_i-r]/\epsilon) +
\exp([r-r_o]/\epsilon)\rbrace,
\end{equation}
where $\epsilon$ is a small parameter characterizing the physical
gaps. We use $\epsilon=6\times 10^{-3}$. Our numerical method
was previously developed to study the end-wall effects in the Taylor-Couette 
system with co- and counter-rotating cylinders~\cite{AGLM08,ADML2012}. 
In the present work we use up to $n_r=50$ and $n_z=500$ Chebyshev
modes in the radial and axial directions, respectively, and up to 
$n_\theta=100$ Fourier modes in the azimuthal direction. The time
step is chosen to be $\delta t=10^{-6}$.

\subsection{Qualitative description of basic dynamical states} 
\label{subsec:BDS}

\subsubsection{Low order instabilities}

In the wide-gap, three-dimensional Taylor-Couette system, various flow
patterns and their bifurcation behaviors are relatively well 
understood~\cite{Coles:1965,ALS86}. In our quasi two-dimensional setting, 
a number of known low-order, non-turbulent instabilities still occur, 
which constitute the background flow upon which a new type of ring 
burst structures emerge. Here we describe these low-order instabilities.
To visualize and distinguish qualitatively different flow patterns,
we use the contour plots of the azimuthal vorticity component $\eta$.

\paragraph*{Primary instability - Taylor vortex flow.} 
As the Reynolds number $Re$ is increased, the basic state, circular
Couette flow (CCF) becomes unstable and is replaced by the classical 
Taylor vortex flow (TVF) that consists of toroidally closed vortices. TVFs 
with some increasing numbers of vortices appear gradually over a large range of 
$Re$, starting from a single Taylor vortex cell generated by the mechanism
of Ekman pumping near the end walls. This initial cell can appear either 
near the top or the bottom lid. In a perfectly circular geometry, initial
Taylor vortex cells can occur simultaneously at both lids, but this is 
less likely in realistic systems due to the inevitable imperfections in 
the system. The onset of initial TVF cells was experimentally found for
$Re$ about~\cite{CSA07} $358$. We find, numerically, that the onset value 
is about $356$. As $Re$ is increased from this value, additional vortices 
appear, which enter the bulk from near the lids until they finally fill 
the whole annulus for $Re$ value about $435$ (experimentally the value 
is about~\cite{CSA07} $437$). For example, we observe a TVF with 22 pairs
of vortices within the annulus, with the characterizing wavenumber of 
$k=3.427$, and an additional pair of Ekman boundary layer vortices near 
the top and bottom lids.

\begin{figure}
\begin{center}
\includegraphics[width=0.95\linewidth]{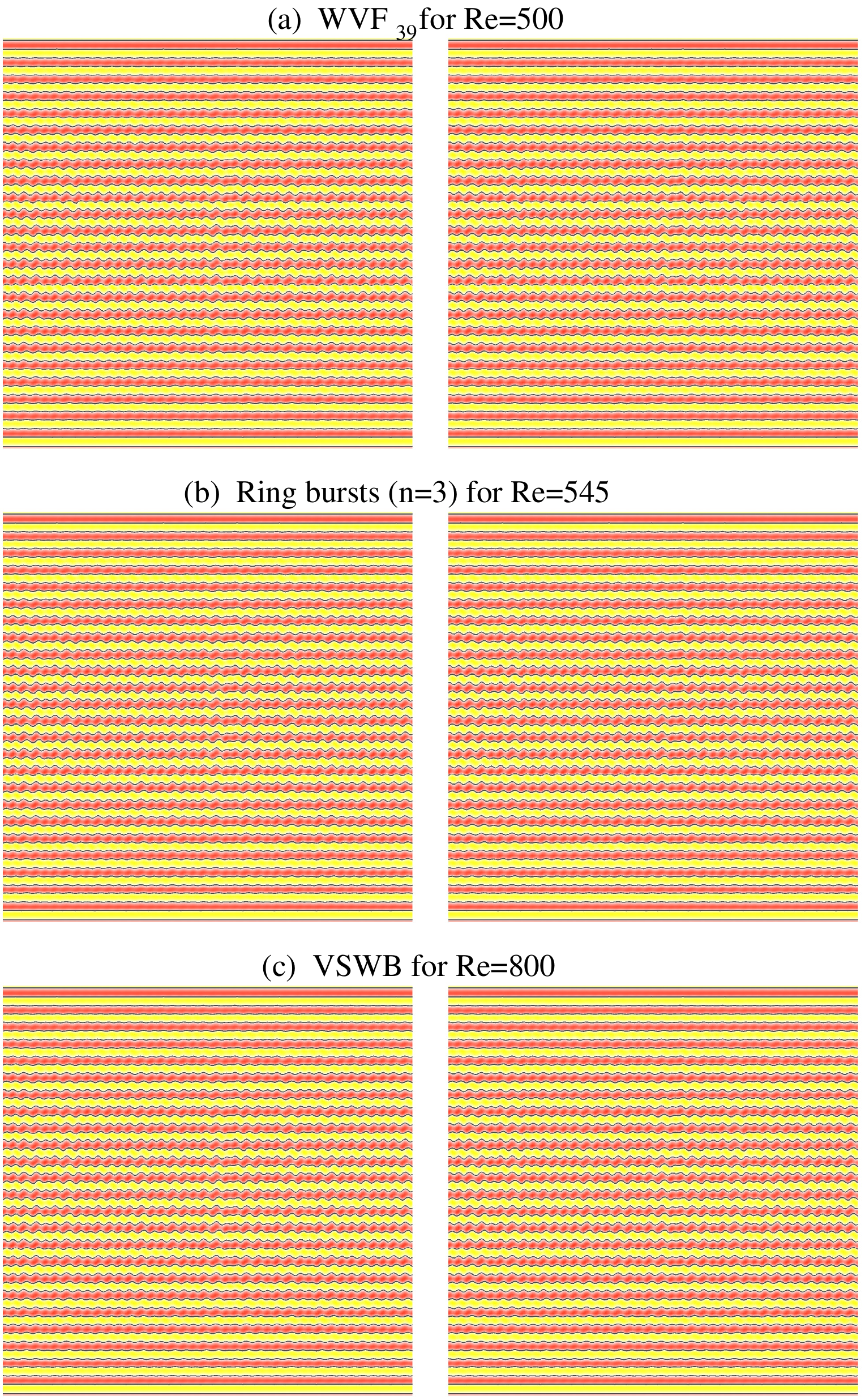}
\end{center}
\caption{(Color online) {\bf Distinct flow patterns in quasi two-dimensional
Taylor-Couette system.} Contours of the azimuthal vorticity component
$\eta$ with (left column, $\eta \in [-400,400]$) and without (right column, 
$\eta(m\neq0) \in [-200,200]$) the axisymmetric component on an 
unrolled cylindrical surface at the midgap ($r=d/2$) for different flows 
and values of $Re$. Red (dark gray) and yellow (light gray) colors 
correspond to positive and negative values, respectively, and the black 
line indicates the zero contour.}
\label{fig:contours_vort_full_without_zero}
\end{figure}

\paragraph*{Secondary instabilities - wavy vortex flow.}
Further increasing $Re$, the TVF becomes unstable and is replaced by     
wavy vortex flow (WVF), a kind of centrifugal instability that appears 
via a supercritical Hopf bifurcation, which exhibits a wavy-like modulation 
but the vortices remain azimuthally closed. We find that the onset of WVF
occurs for $Re \approx 473$ (experimentally the value is about~\cite{CSA07}
$475$). The WVF, as shown in Fig.~\ref{fig:contours_vort_full_without_zero}$(a)$,
is associated with relatively {\em high} values of the azimuthal wave number 
$m$. In general, WVF states consisting of many azimuthal modes 
can occur. For example, the WVF pattern in 
Fig~\ref{fig:contours_vort_full_without_zero}$(a)$ contains $m=39$ modes,
which we name as WVF$_\text{39}$. We find that this particular WVF state is 
in fact a global background flow state in a wide range of $Re$ values.

The TVF and WVF patterns occur regardless of whether 
the flow is three dimensional or quasi two-dimensional, although the 
critical values of $Re$ for the onsets of these flow patterns depend 
on the system details.  

\subsubsection{High-order instabilities}

\paragraph*{Very short wavelength bursts and localized patches.}
For the commonly studied~\cite{ALS86}, wide-gap, three-dimensional 
Taylor-Couette system, e.g., of radius ratio $0.883$, the typical 
sequence of solutions with increasing $Re$ values is as follows. 
As the stationary TVF becomes unstable, time dependent WVF arises, 
followed by the so-called modulated WVF~\cite{ALS86}, eventually 
leading to turbulent behaviors. However, in the narrow-gap, quasi
two-dimensional system, we have not observed the global transition 
from WVF to modulated WVF. Instead, we find that very short wavelength
bursts (VSWBs) appear directly after the onset of WVF {\em without} 
any other types of intermediate solutions. Onset of VSWBs is
$Re \approx 483$, which agrees with the experimentally observed onset
value~\cite{CSA07}. 

As $Re$ is increased, we observe {\em localized patches} (LPs) 
superposed on WVF. The 
pattern within LPs can be either wavy-like or turbulent, depending
on the $Re$ value (lower and high values for the former and the latter,
respectively). In particular, LPs with wavy-like interior behaviors emerge,
and for $Re > 580$ LPs with turbulent behaviors arise, which can either
grow to VSBWs or decay slowly. The number of LPs depends on $Re$ as well. 
In general, the higher the value of $Re$ the larger the number of LPs appears.  
The patches are randomly distributed over the whole bulk length and are always 
visible except for the region near the lids in which strong Ekman vortices 
arise. 

\begin{figure}
\begin{center}
\includegraphics[width=0.9\linewidth,height=17cm]{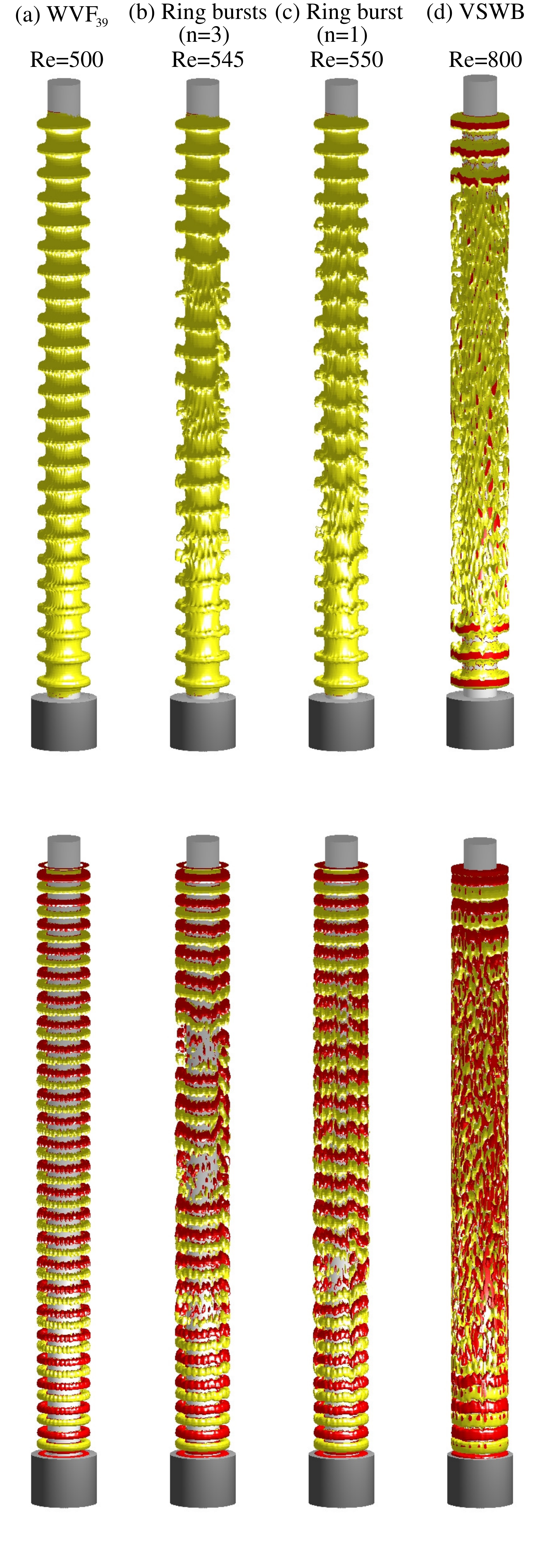}
\end{center}
\caption{(Color online) {\bf Angular momentum and vorticity.} 
Isosurfaces of $rv$ (top row) and $\eta$ (bottom row)
for flows at different values of $Re$ (isolevels shown for the
top and bottom rows are $rv=80$ and $\eta = \pm30$, respectively). 
For clear visualization, here and in all following 
three-dimensional plots the radius ratio is scaled by the factor 100.}
\label{fig:isosurf_rv_eta_G44}
\end{figure}

\paragraph*{Ring-burst patterns.}
When $Re$ is increased above about $540$, we discover a {\em new} type of 
localized, intermittent burst solution, the {\em ring bursts} that coexist 
with VSWBs. While both types of solutions are localized, there are 
characteristic differences. For example, VSWBs appear randomly over the 
whole bulk fluid region with seemingly expanding behaviors in all 
directions, but ring bursts always remain {\em localized} in the axial 
direction. In fact, the bursts are generated from the localized turbulent
patches that grow in the azimuthal direction as $Re$ is increased. For  
sufficiently high values of $Re$, the patches extend over one circumference,
generating distinct ring bursts that are separated from the flow patterns
in the rest of the bulk, as shown in 
Fig.~\ref{fig:contours_vort_full_without_zero}(c). The characteristics of
the background flows in the regions surrounding the ring bursts depend 
strongly on $Re$. They can range from wavy-like patterns (i.e., WVF$_{39}$) 
to inter-penetrating spirals due to the interactions among various azimuthal 
modes, as shown in Fig.~\ref{fig:contours_vort_full_without_zero}(c).
We observe ring burst patterns of different order $n$,
as shown in Figs.~\ref{fig:contours_vort_full_without_zero} 
and \ref{fig:isosurf_rv_eta_G44}). For example, we observe $n \in \{1,2,3\}$. 
All these states coexist but the probability for ring bursts with larger
values of $n$ increases with $Re$. Along the axial direction, the ring
burst patterns can appear at any position, except for the regions of Ekman 
vortices near the lids, because strong boundary layer vortices prevent 
the development of ring burst patterns.

While there is no apparent order associated with the flow patterns within 
the ring bursts, the flow in the regions in between exhibit a clear signature of
WVF$_{39}$ pattern. The ring bursts can spontaneously break up and 
disappear. Depending on the number of ring bursts, the transient time 
that it takes for the burst to decay into localized patches can be 
relatively long. In addition, there can be transitions between patterns
with distinct numbers of ring bursts. For example, suppose there is a 
ring-burst pattern of order $n = 3$. If one ring disappears,
a new ring-burst region can appear and grow. Transitions between patterns
with either increasing or decreasing numbers of ring bursts have been
observed.

For $Re > 850$, we find that VSWBs fill the whole bulk fluid region, as
shown in Fig.~\ref{fig:contours_vort_full_without_zero}(c). In fact, 
for $Re > 900$, the whole fluid region is saturated with VSWBs. These
numerical observations are in agreement with experimental findings~\cite{CSA07}. 
Note, however, even when VSWBs fill the interior of the bulk WVF-like 
patterns always appear near the axial end-walls due to the fact that 
Ekman vortices can stabilize the boundary layers against turbulent bursts. 
The size of the WVF-like Ekman regions depends on $Re$ - for higher value 
of $Re$ the smaller the WVF regions near the lids appear, as can be seen from
Fig.~\ref{fig:contours_vort_full_without_zero}.

\paragraph*{Underlying flow patterns.}
The right panels in Fig.~\ref{fig:contours_vort_full_without_zero} show 
contours for the same pattern as those for the left panels but {\em without} 
the underlying axisymmetric contribution. The resulting ``reduced'' flow 
pattern of WVF$_{39}$ in Fig.~\ref{fig:contours_vort_full_without_zero}(a) 
appears quite regular, indicating the existence of the dominant wavy mode 
with $m=39$ in the azimuthal direction. The black curves in the vertical 
direction represent the zero contour lines. At several azimuthal positions,
these lines narrow, indicating the emergence of a second but weaker mode, 
e.g., $m=8$. For order 3 ring bursts, the flow patterns near the lids are 
somewhat modified due to the presence of higher $m$ modes, as shown in  
Fig.~\ref{fig:contours_vort_full_without_zero}(b). The flow patterns in 
the central region (including that containing the three ring bursts) exhibit 
a completely different behavior. In particular, with respect to contours 
$\eta(m\neq0)$ the turbulent ring bursts and the separating WVF$_{39}$ 
pattern appear {\em indistinguishable}. This finding suggest a quite strong 
axisymmetric dominance of the flow even when ring bursts arise,
confirming that the bursting structure is ordered in azimuthally closed 
rings around the cylinder with the same $m=0$ axisymmetry. For turbulent flows 
or VSWBs in Fig.~\ref{fig:contours_vort_full_without_zero}(c), none of the 
patterns has such an ordered structure.

\section{Characterization of ring bursts} \label{sec:RB_characterization} 

\subsection{Angular momentum, azimuthal vorticity, and modal kinetic energy} 
\label{subsec:AMKE}

Figure~\ref{fig:isosurf_rv_eta_G44} shows the isosurfaces of the 
angular momentum $rv$ (top row) and azimuthal vorticity $\eta$ 
(bottom row) for representative flow patterns at different values of
$Re$. The dominant axisymmetric contribution and high azimuthal wavenumber
associated with WVF$_{39}$, as shown in Fig.~\ref{fig:isosurf_rv_eta_G44}(a),
serve as the background pattern for ring bursts and VSWBs. The three turbulent,
azimuthally closed burst regions associated with the order-3 ring burst pattern
is distinctly visible, as shown in Fig.~\ref{fig:isosurf_rv_eta_G44}(b).
Within each burst region, both $rv$ and $\eta$ appear random but the 
(background) flow patterns in between the bursting regions are remnant
of the WVF$_{39}$ pattern. With increasing $Re$ higher $m$ modes emerge,
but the separation between burst and non-burst regions persists. For  
high values of $Re$, VSWBs arise, as shown in Fig.~\ref{fig:isosurf_rv_eta_G44}(d).
In this case, isosurface plots of $rv$ and $\eta$ exhibit random flow patterns 
except for the regions near the lids due to the Ekman boundary layers.

\begin{figure}
\begin{center}
\includegraphics[width=0.6\linewidth]{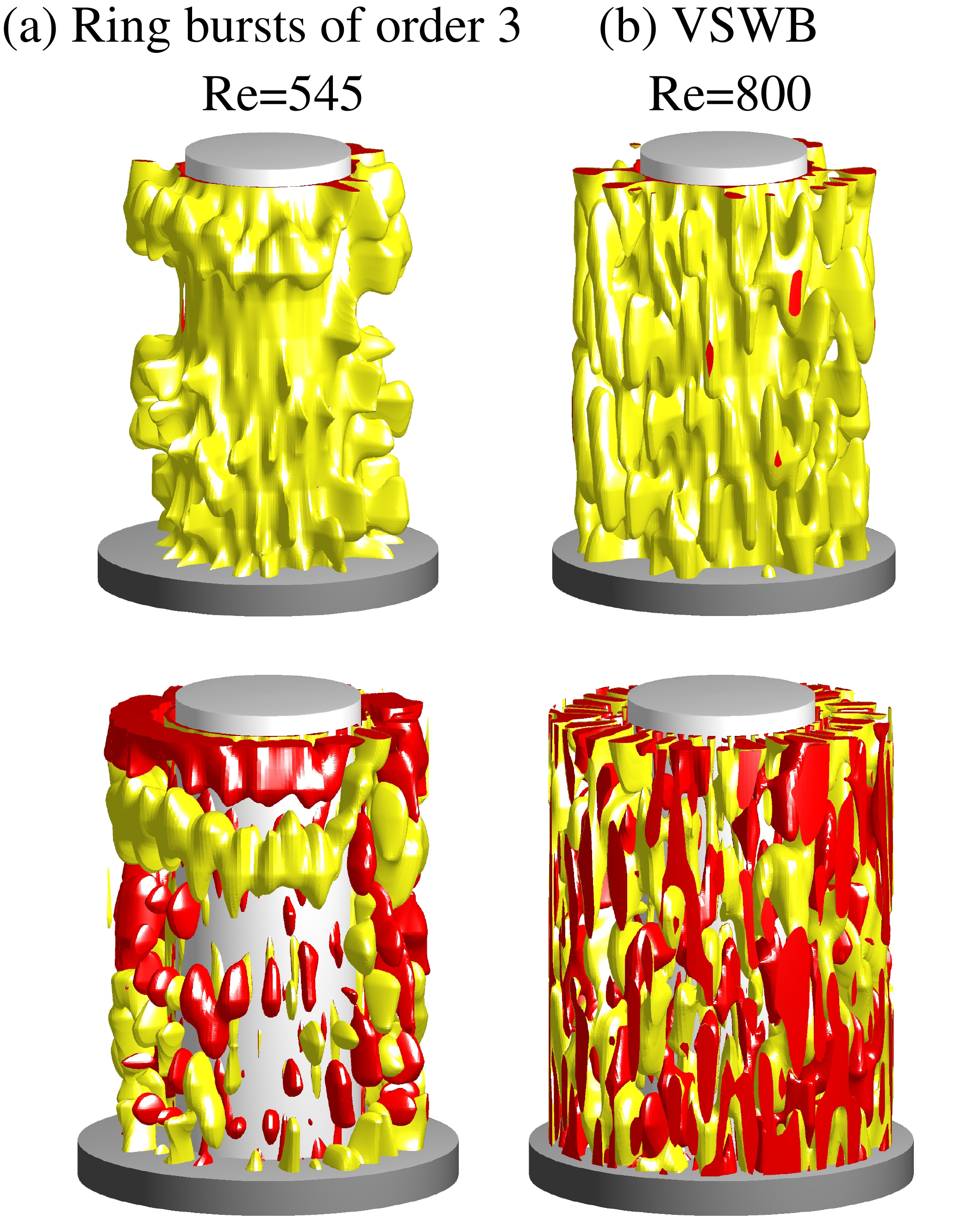}
\end{center}
\caption{(Color online) {\bf Three-dimensional views of angular momentum
and vorticity.} Magnitude of the angular momentum (top panels)
and azimuthal vorticity (bottom panels) for a flow segment of vertical 
length $\Gamma/10$ for ring bursts (left column) and VSWB (right column). 
The segment is taken from the region containing the middle ring burst in   
Fig. \ref{fig:isosurf_rv_eta_G44}(b).}
\label{fig:isosurf_rv_eta_cut-out_G4}
\end{figure}

To better visualize and illustrate the similarity and differences between 
ring burst and VSWB patterns, we present in 
Fig.~\ref{fig:isosurf_rv_eta_cut-out_G4} the behaviors of the angular 
momentum (top panels) and azimuthal vorticity (bottom panels) for a 
segment of the bulk flow. The length of the segment is
$\Gamma/10$. We see that, for the ring burst pattern (left column),
there is a dominant axisymmetric component, but this is lost for VSWB. 
While the entire structure of the ring-burst pattern possesses an 
axisymmetry due to its azimuthal dominance ($m=0$), no apparent 
symmetry exists in the interior of the burst regions. There are thus
two distinct spatial scales associated with ring bursts: a large scale
determined by the axisymmetry and a small scale present in the interior
of the bursting regions. 

\begin{figure}
\begin{center}
\includegraphics[width=0.95\linewidth]{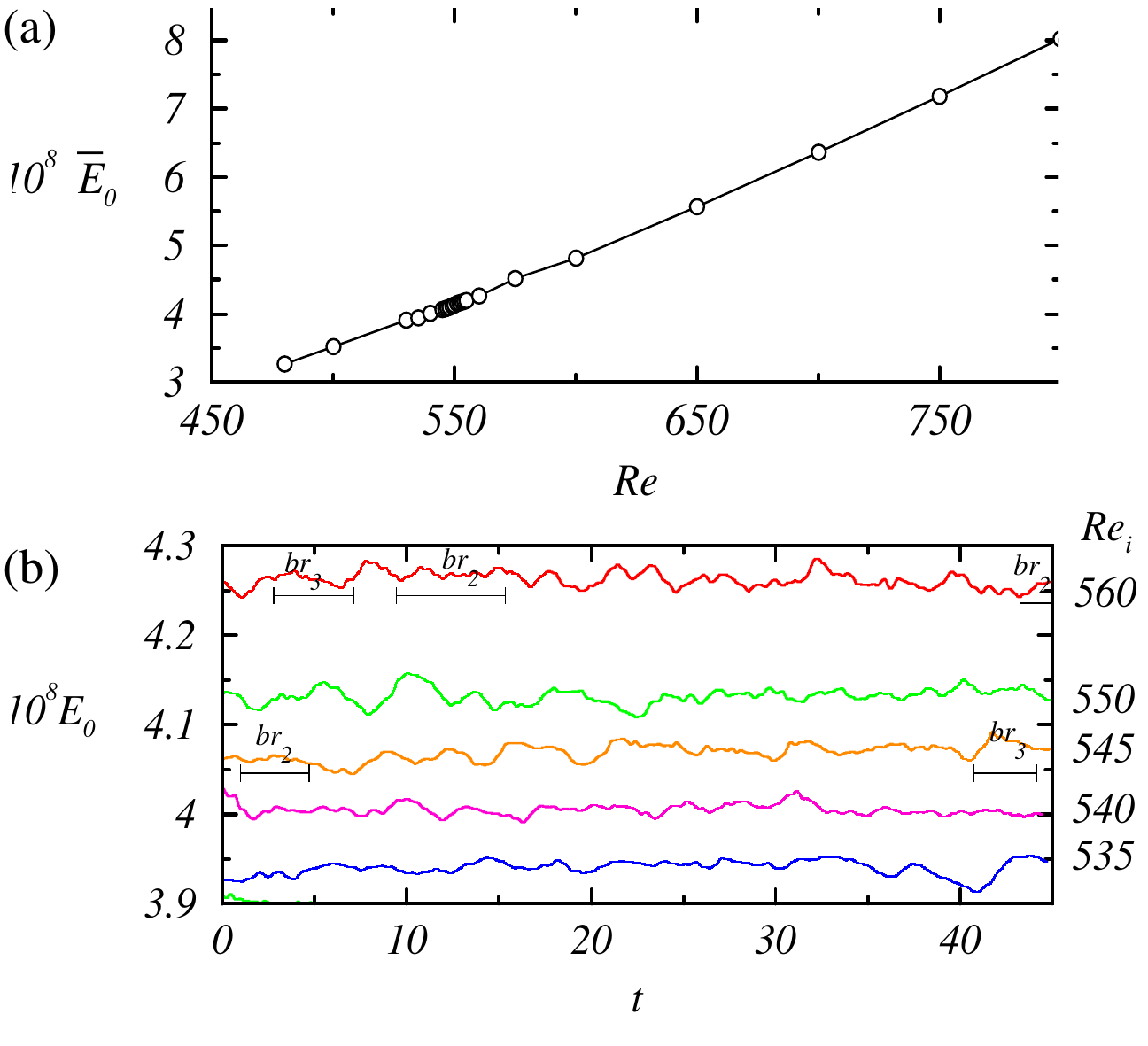}
\end{center}
\caption{(Color online) {\bf Behaviors of the kinetic energy.}
(a) Time-averaged ($\approx$ 50 diffusion time) axisymmetric energy 
$\overline{E}_0$ versus $Re$ and (b) time series of $E_0$ for 
a number of representative values of $Re$. The time intervals in 
which distinct ring-burst patterns last are indicated. For example,
for $Re=560$, ring burst pattern of order 3 is present for $2.5<t<6.9$,
and that of order 2 can be found for $9.2\lesssim t\lesssim15.8$ and 
$42.6\lesssim t\lesssim47.1$. For $Re=545$, order-2 ring burst pattern
appears for $0.5\lesssim t\lesssim 4.5$ and the order-3 pattern is present
for $40.3\lesssim t\lesssim44$. However, the ring burst patterns are 
not reflected in $\overline{E}_0(t)$.}
\label{fig:E0-Re-t}
\end{figure}

We next examine the modal kinetic energy defined as  
\begin{equation} \label{Eq_energy}
E := \sum_{m} E_m = \int_0^{2\pi} \int_{0}^{\Gamma} \int_{r_i}^{r_o}
{\bf u}_m {\bf u^*}_m r \textrm{d}r \textrm{d}z \textrm{d}\theta ,
\end{equation}
where ${\bf u}_m$ (${\bf u^*}_m$) is the $m$-th (complex conjugated) Fourier 
mode of the velocity field. Due to time dependence of the solutions it is 
necessary to calculate the time averaged kinetic energy $\overline{E}$. 
In the energy characterization local quantities such as the radial velocity 
at mid-height and mid-gap, $u_r(d/2,0,\Gamma/2,t)$, are relevant.

Figures~\ref{fig:E0-Re-t}(a) and \ref{fig:E0-Re-t}(b) show the 
time-averaged axisymmetric energy $\overline{E}_0$ versus $Re$ and 
its time evolutions for a number of representative $Re$ values, respectively.
We see that $\overline{E}_0$ increases monotonously with $Re$, regardless
of the nature of the bulk flow patterns. Likewise the time evolution of the 
kinetic energy presents no clear indication of ring burst patterns.
For example, for $Re=545$, in the time interval $[40,44]$ the flow pattern 
should be order-3 ring burst, but the energy time evolution does not 
show any signature of this burst pattern.

\begin{figure}[htbp!]
\begin{center}
\includegraphics[width=0.7\linewidth]{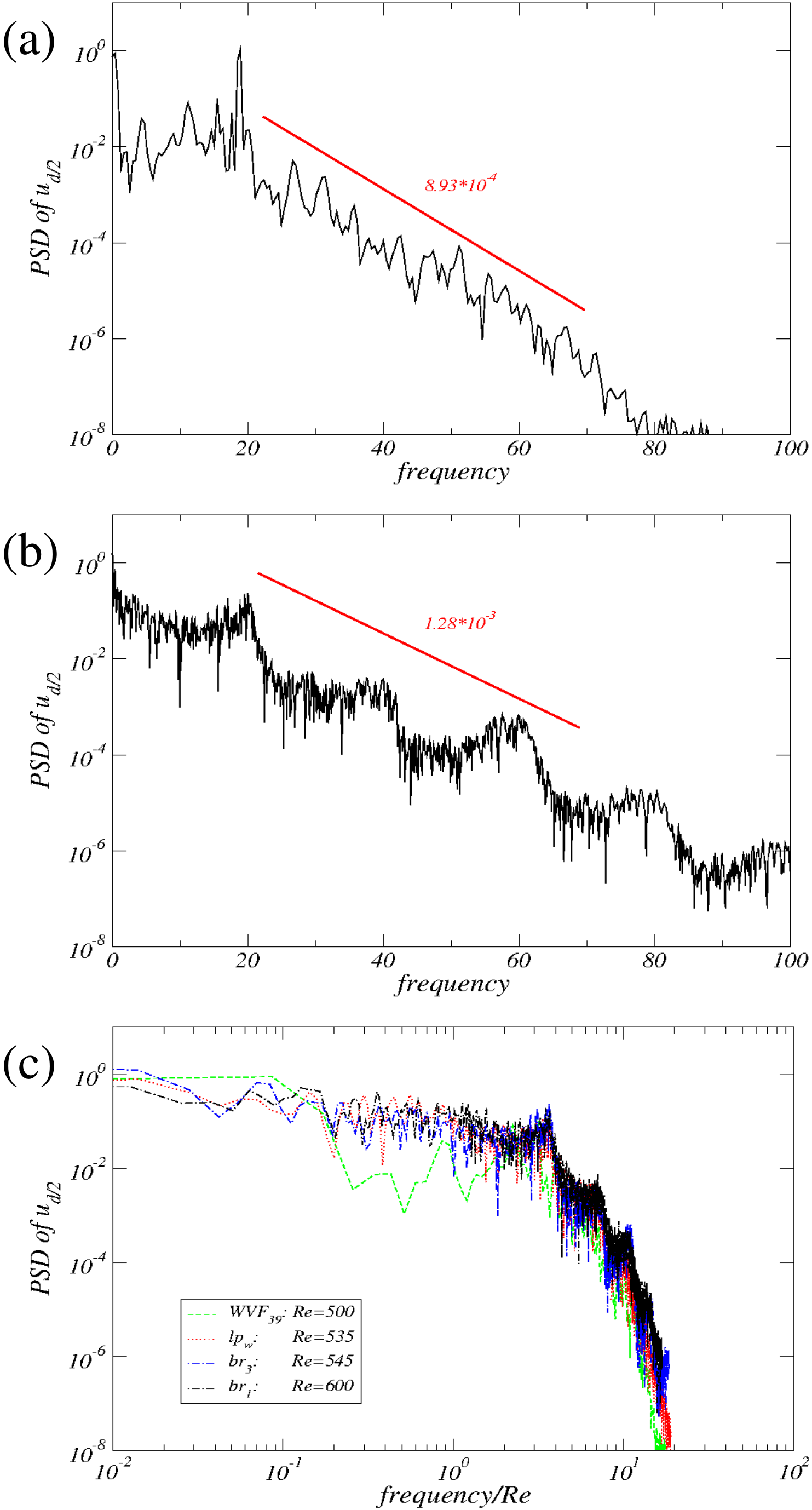}
\end{center}
\caption{(Color online) {\bf Power spectral density (PSD).} 
PSD curves calculated from the radial flow component 
$u_{d/2}=u(d/2,0,\Gamma/2,t)$ for (a) WVF$_{39}$ ($Re=500$) 
and (b) ring burst of order 3 for $Re=545$. (c) PSD curves scaled 
by the respective Reynolds number for different flow patterns. As the number 
of ring bursts is increased, the local peaks in the PSD curves
become more pronounced.}
\label{fig:u-t_PSD_u-t_Re-var}
\end{figure}

Figures~\ref{fig:u-t_PSD_u-t_Re-var}(a) and \ref{fig:u-t_PSD_u-t_Re-var}(b) 
show the power spectral density (PSD) of the radial velocity profile at the midgap 
for WVF and ring burst, respectively. We see that the PSD associated with 
the ring burst pattern (b) indicates the existence of significantly higher 
modes than the WVF pattern (a). In fact, the PSD of WVF$_{39}$ in (a) shows 
a strong peak at the frequency of about 19, which corresponds to the dominant 
azimuthal mode ($m=39$). This peak is still present in (b) but it is much
broadened, indicating WVF as the background flow pattern for the ring burst.
Figure~\ref{fig:u-t_PSD_u-t_Re-var}(c) shows the scaled PSD curves for several 
flow patterns for different values of $Re$. We see that the PSD curves for the 
ring-burst patterns essentially collapse into one frequency band, but the PSD 
curve associated with WVF$_{39}$ lies slightly below those of ring burst 
patterns. There is relatively large difference for small frequencies but 
it becomes insignificant for higher frequencies. This is
further support for the role of WVF$_{39}$ in providing the skeleton 
structure for all ring burst patterns.

\subsection{Cross-flow energy} \label{subsec:crossflow}

\begin{figure}
\begin{center}
\includegraphics[width=0.95\linewidth]{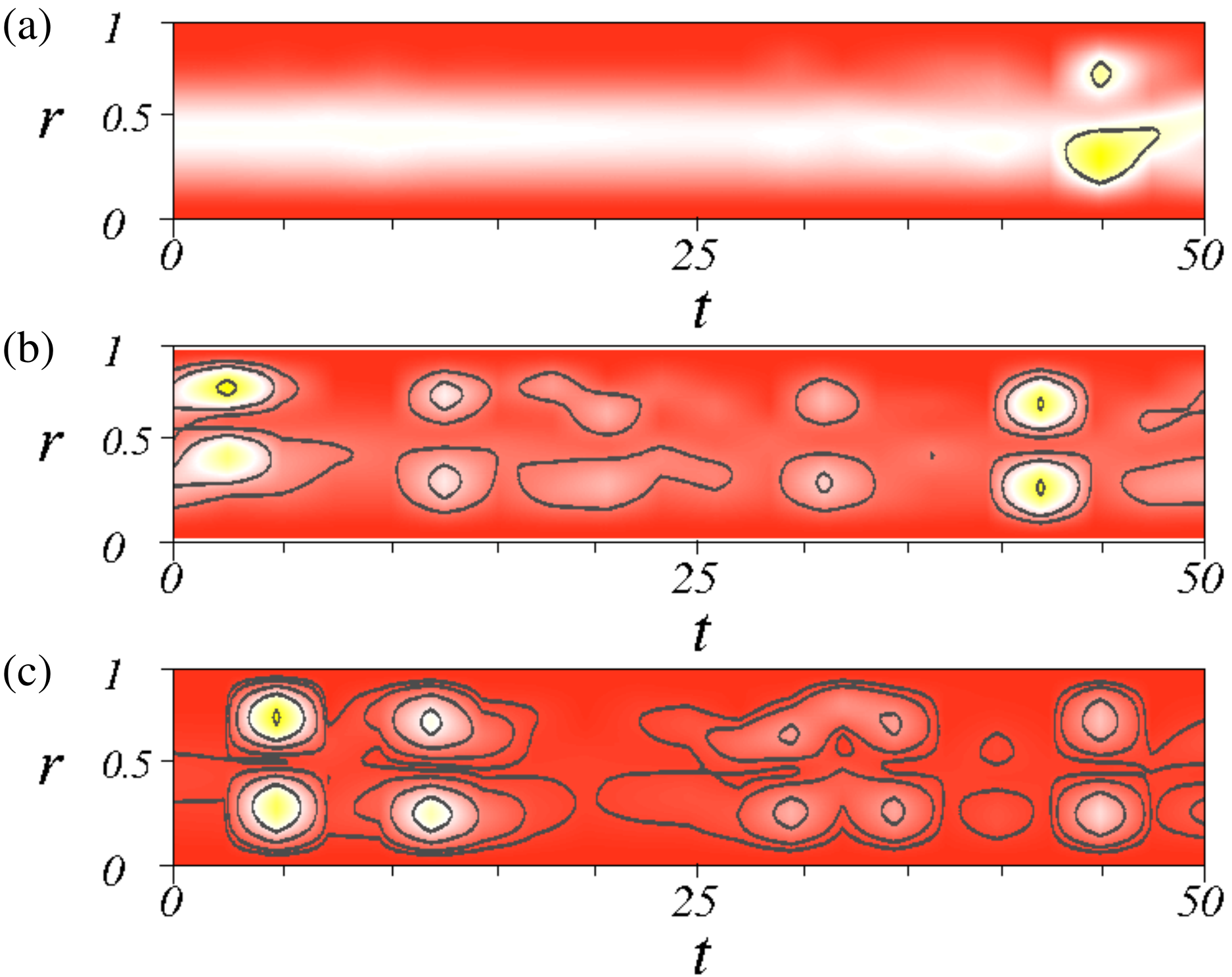}
\end{center}
\caption{(Color online) {\bf Radial component of cross-flow energy.} 
Spacetime plots of the radial component of the cross-flow energy, 
$E^{cf,r}(r,t) = \langle u_r^2 + u_z^2 \rangle_{A(r)}$, averaged 
over the surfaces $A$ of a concentric cylinder at radius $r$ for 
(a) $Re = 500$, (b) $Re = 545$, and (c) $Re = 560$. Red (yellow) 
color indicates high (low) energy with contours defined as 
$\Delta E^{cf,r} = 5\times 10^5$. The maximum energy values in (a-c)
are approximately $8.1 \times 10^5$, $8.5 \times 10^6$, and 
$1.2 \times 10^6$, respectively. For better visualization the  
radial gap width is magnified by a factor of 500 (the same for
Fig.~\ref{fig:crossflow-t} below).}
\label{fig:crossflow_all}
\end{figure}

From Fig.~\ref{fig:E0-Re-t}(b), we see that the modal kinetic 
energy is not effective at distinguishing ring bursts from other basic
flow patterns. A suitable and commonly used quantity in the study of fluid
turbulence is the so-called cross-flow energy, where typically small values 
indicate a flow being laminar, large values corresponding to turbulence.

\subsubsection{Radial component of cross-flow energy}
\label{subsubsec:radial_component}

The radial component of the cross-flow energy is given by~\cite{BrEc2013}
\begin{equation} \label{EQ:crossflow}
E^{cf,r}(r,t) = \langle u_r^2 + u_z^2 \rangle_{A(r)},
\end{equation}
where $\langle\cdot\rangle_{A(r)}$ denotes averaging over the surface of
a concentric cylinder at radius $r$. The cross-flow energy component
$E^{cf,r}(r,t)$ measures the instantaneous energy associated with the 
radial and axial velocity components at radial distance $r$. 
Figure~\ref{fig:crossflow_all} shows the spacetime plots of $E^{cf,r}(r,t)$ 
over the time period of 50 DT for three values of $Re$. We observe the 
temporal emergence and disappearance of various ring-burst patterns in the
bulk. For example, for $Re=545$, the order-3 ring burst pattern exists 
for $40 \leq t \leq 44$, and the order-2 pattern appears for 
$0.5 \leq t \leq 4.5$, as shown in Fig.~\ref{fig:crossflow_all}(b). 
For $Re=560$, an order-3 pattern appears for $2.5 \leq t \leq 7.0$,
and a single ring burst pattern (order 1) exists for $43 \leq t \leq 47$, 
as shown in Fig.~\ref{fig:crossflow_all}(c). 

Plots of the cross-flow energy exhibit two features. Firstly, presence 
of the order-1 ring burst pattern is accompanied by a significant increase 
in the radial cross-flow energy as compared with that associated with the 
background flow, e.g., WVF$_{39}$, as indicated by the uniform red (dark 
gray) regions. Secondly, the profile of $E^{cf,r}(r,t)$ for any ring burst 
pattern is approximately symmetric with respect to the middle of 
the gap. It is thus plausible that the ring-burst patterns are  
similar to the so-called {\em turbulent WVF} 
state~\cite{TsFe2004,OMMSK1998,Don08a,Don2007} 
(see Sec. \ref{SUBSEC:axial_space}). The symmetry may be a consequence of 
the quasi two-dimensional nature of the flow.
In all cases, the occurrence of ring burst patterns enhances the    
magnitude of the cross-flow energy.   

\begin{figure}
\begin{center}
\includegraphics[width=0.8\linewidth]{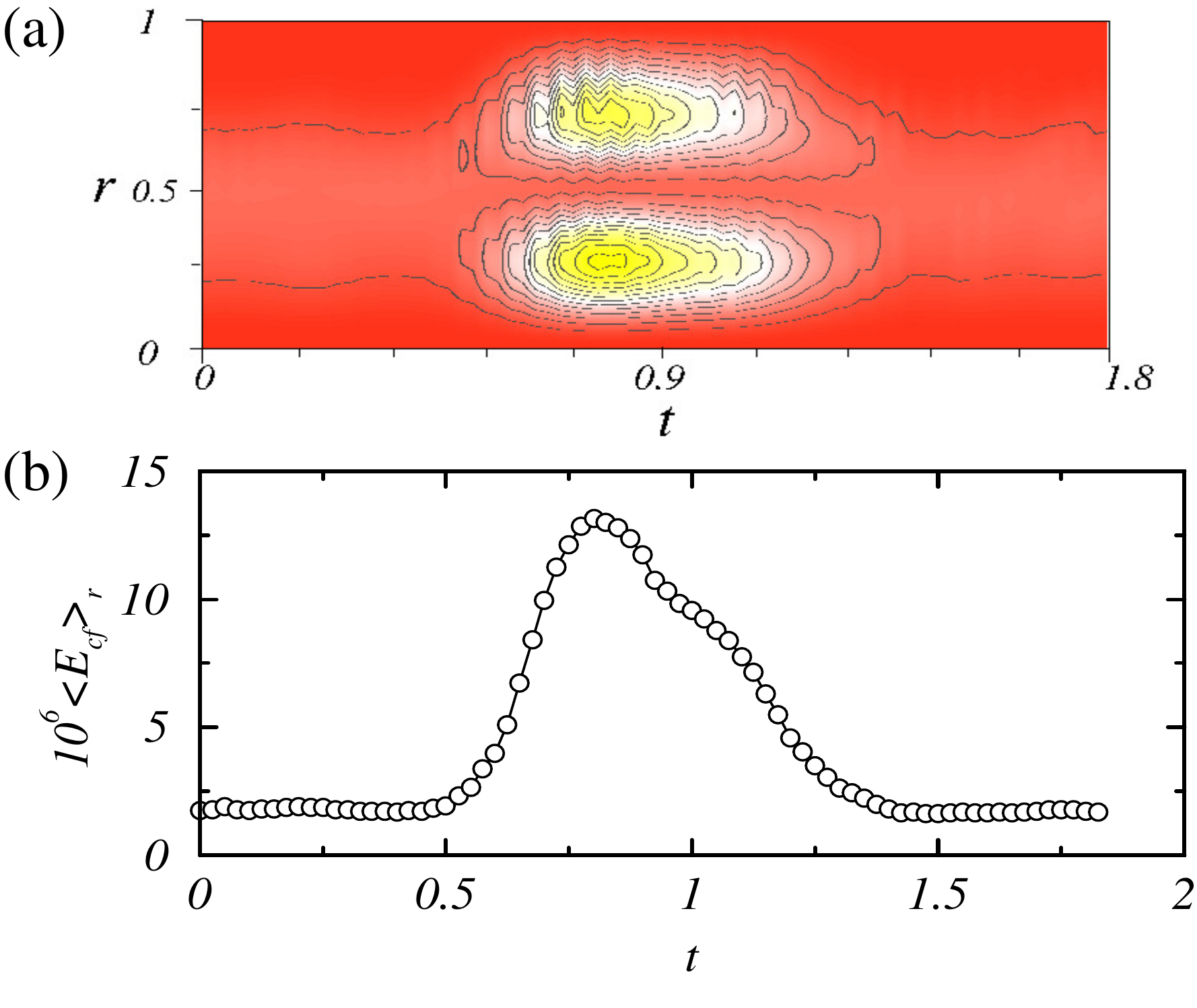}\
\end{center}
\caption{(Color online) {\bf Magnified view of the radial component of the
cross-flow energy.} (a) Magnification of a segment of 
Fig.~\ref{fig:crossflow_all}(b) for $t\in[0,1.8]$ illustrating the emergence
and disappearance of order-3 ring burst pattern for $Re = 545$. The contours
are defined through $\Delta E^{cf,r} = 5\ast 10^5$, and the maximum energy 
value is about $8.288 \times 10^6$. (b) Averaged radial cross-flow energy 
$\langle E^{cf,r} \rangle_r$ versus $t$ for order-3 ring burst for $Re = 545$.
The nearly constant background flow is WVF$_{39}$ with 
$\langle E^{cf,r} (\text{WVF}_{39})\rangle_r \approx 2 \times 10^{-6}$.}
\label{fig:crossflow-t}
\end{figure}

Figure~\ref{fig:crossflow-t}(a) shows a magnification of a segment of 
Fig.~\ref{fig:crossflow_all}(b) for $t\in[0,1.8]$, which contains the
emergence and disappearance of order-3 ring burst for $Re = 545$, 
occurring at $t \approx 0.45$ and $t \approx 1.35$, respectively. Outside 
this time interval the flow is WVF$_{39}$, which constitutes a nearly
uniform background without any significant variation in $E^{cf,r}(r,t)$.
The spatial distribution of $E^{cf,r}(r,t)$ exhibits a kind of symmetry
about $r = 0.5$. Figure~\ref{fig:crossflow-t}(b) shows the average 
cross-flow energy, $\langle E^{cf,r}\rangle_r$, as a function of time $t$,
corresponding to the emergence and disappearance of the ring burst pattern.
We observe a significant enhancement of the average radial energy over that 
of the background flow ($\langle E^{cf,r}\rangle_r\approx 2\times 10^{-6}$.
In fact, in the time interval where the ring burst exists, the maximum 
value of the average radial energy is about one order of magnitude 
larger than that of the background flow. In general, the maximum energy
depends on the number $n$ (order) of ring bursts in the annulus, where
a larger value of $n$ corresponds to larger value of the maximum energy.

\begin{figure}
\begin{center}
\includegraphics[width=0.8\linewidth,height=8cm]{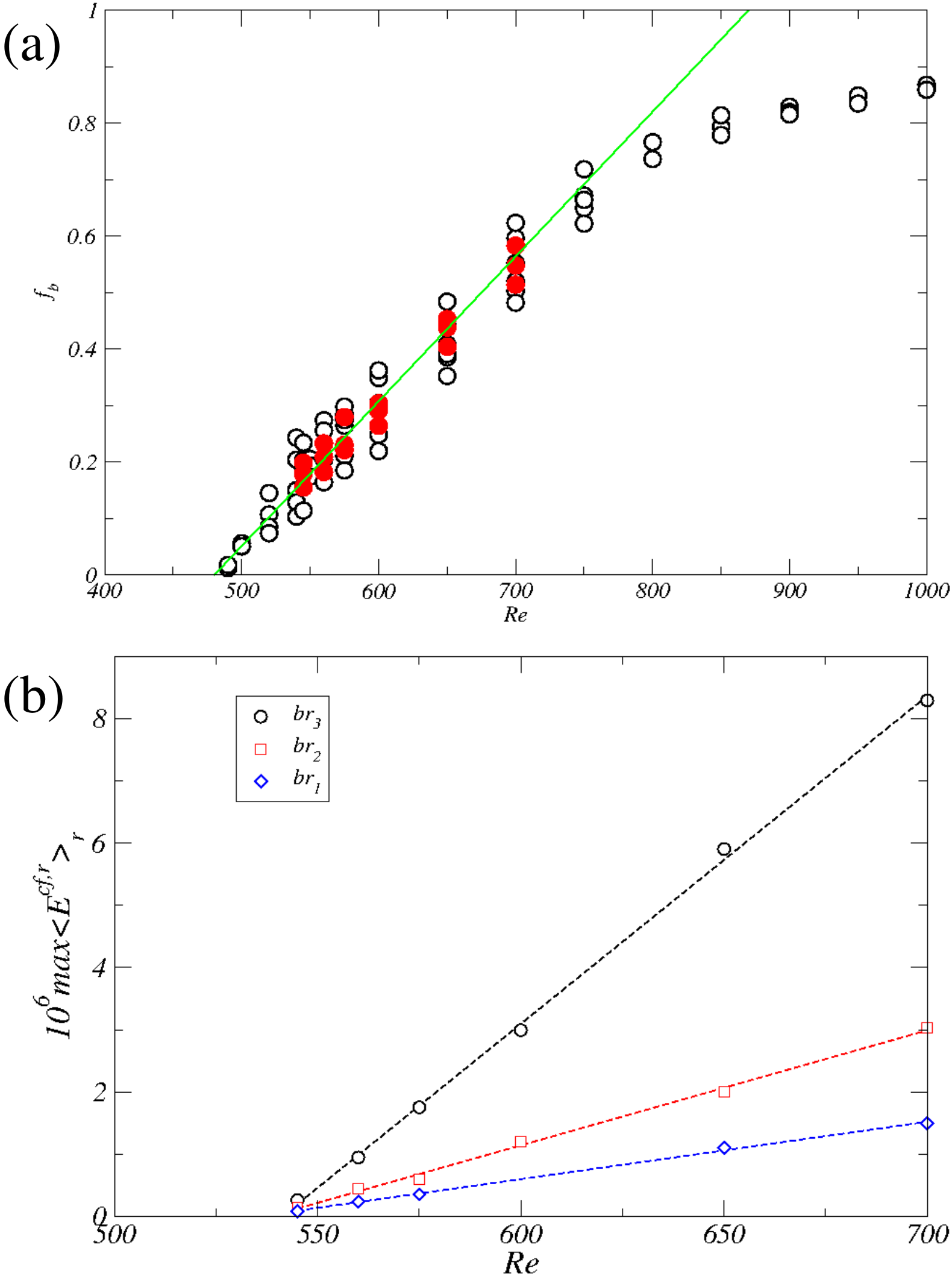}
\end{center}
\caption{(Color online) {\bf Onset of burst patterns.}
(a) Burst fraction $f_b$ as a function of $Re$. A linear fit indicates 
that onset of VSWBs occurs for $Re \approx 480$ (the experimental 
onset~\cite{CSA07} is $Re \approx 483$). (b) The maximum average radial
energy $\langle E^{cf,r}\rangle_r$ [Eq.~\eqref{EQ:crossflow}] versus
$Re$ for ring bursts of different order ($n\in\{1,2,3\}$).
The linear behavior suggests that ring bursts are result of a forward 
bifurcation, the onset of which occurs for $Re_c \approx 537$. In this
figure, the average energy of the underlaying background flow, 
$\langle E^{cf,r}\rangle_r$, has been subtracted off.}
\label{fig:max_crossflow-Re}
\end{figure}

The emergence and development of any type of burst patterns with 
increasing $Re$ can be conveniently characterized using the 
quantity $f_B$, the percentage of the annulus showing bursts in 
the spacetime plot. Figure~\ref{fig:max_crossflow-Re}(a) shows $f_B$ 
versus $Re$, where we observe approximately a linear behavior for 
$Re \alt 900$, and the transition to bursts occurs for $Re \approx 480$, 
in agreement with the onset of VSWB~\cite{CSA07}. Onset of ring bursts 
can be revealed by examining the maximum value of the average radial
energy, $\langle E^{cf,r}\rangle_r$, versus $Re$, as shown in 
Fig.~\ref{fig:max_crossflow-Re}(b). Regardless of the order $n$ of 
ring bursts, there is a linear increase in $\langle E^{cf,r}\rangle_r$
with $Re$, suggesting a type of forward bifurcation. Calculations of 
the flow amplitudes show a square-root type of scaling behavior with 
increasing parameter difference from the ``critical'' point, providing
further support for the forward nature of the bifurcation. Due to strong 
localization of the ring bursts, the slope of the linear scaling behavior 
depends on the ring-burst order $n$. Nonetheless, the onset value 
$Re_c \approx 537$ of ring bursts does not depend on $n$. This 
is evidence that ring bursts emerge {\em after} VSWBs.

\subsubsection{Axial component of cross-flow energy}
\label{subsubsec:axial_component}

\begin{figure}
\begin{center}
\includegraphics[width=0.8\linewidth,height=8cm]{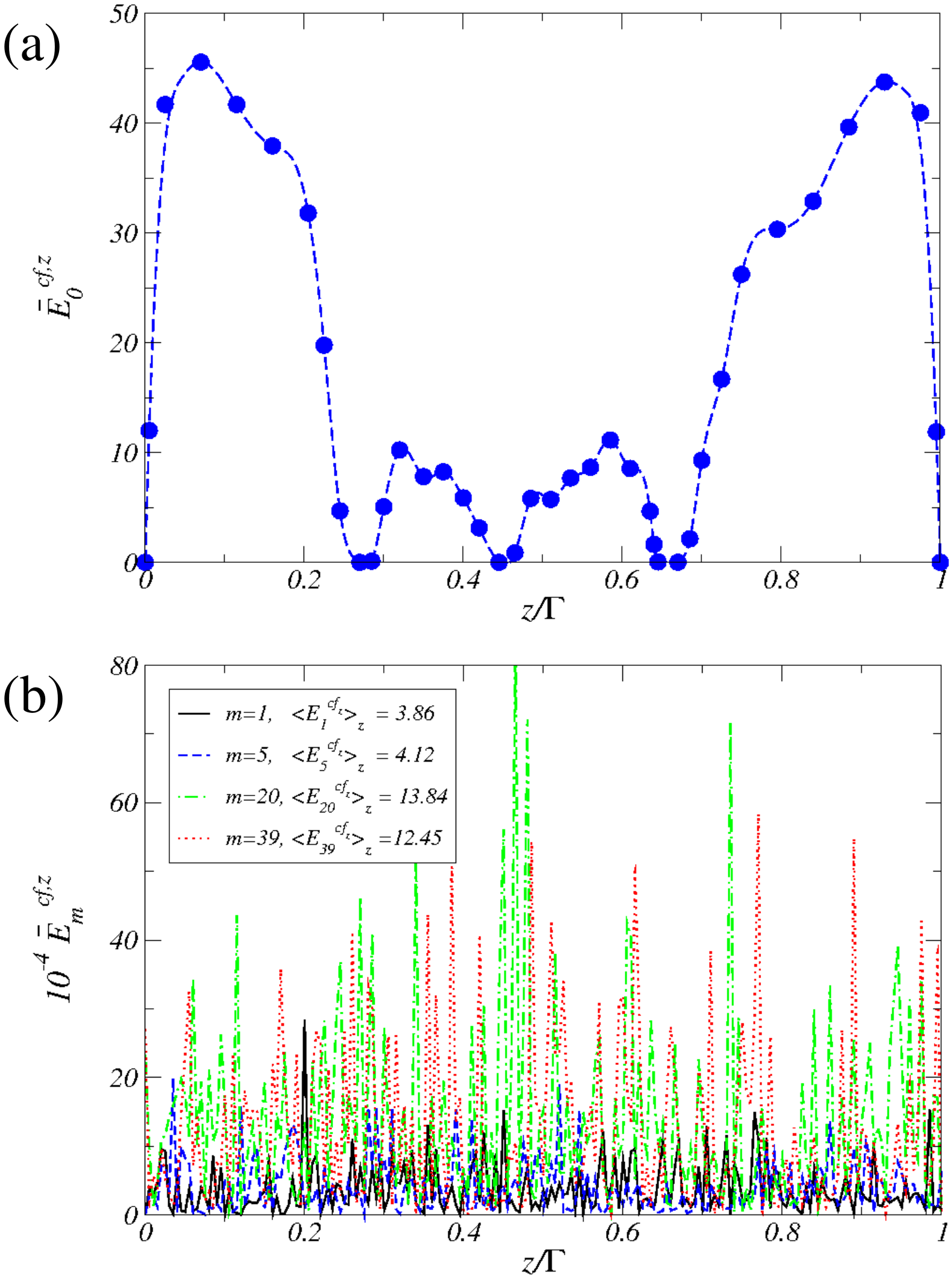}
\end{center}
\caption{(Color online) {\bf Axial component of cross-flow energy.} 
Associated with the order-3 ring burst pattern for $Re = 545$, 
(a) time averaged axisymmetric energy component $\overline{E}^{cf,z}_0$ 
and (b) energy contributions $\overline{E}^{cf,z}_m$ from selected 
higher azimuthal $m$ modes. In (b), additional axial average energy values 
are indicated.}
\label{fig:Ecf-z}
\end{figure}

The axial component of the cross-flow energy is
\begin{equation} \label{EQ:axial_crossflow}
E_m^{cf,z}(z,t) = \langle (u_r)_m^2 + (u_z)_m^2\rangle_{A(z)} . 
\end{equation}
where $A(z)$ stands for averaging over the radial and azimuthal variables
on the surface of a disc at a fixed axial position $z$. Figure~\ref{fig:Ecf-z} 
shows the time-averaged value $\overline{E}_m^{cf,z}$ for the order-3 ring
burst pattern for different azimuthal modes. The axisymmetric component 
$\overline{E}_0^{cf,z}$ is dominant. Near the center of the ring burst 
region in the axial direction, the values of $\overline{E}^{cf,z}_0$ are 
smaller than those around the edges. However, the contributions from all 
higher modes, i.e., $\overline{E}_m^{cf,z}$ for $m > 0$, have similarly 
small magnitudes as compared with $\overline{E}_0^{cf,z}$, as shown in 
Fig.~\ref{fig:Ecf-z}(b). In fact, for higher $m$ modes, the axial 
components of their cross-flow energies are randomly distributed over
$z$ and they are not indicators of any appreciable difference between 
the background flow and the ring burst pattern.

\subsubsection{Mode separation of axial cross-flow energy} 
\label{subsubsec:mode_separation}

\begin{figure}
\begin{center}
\includegraphics[width=0.9\linewidth,height=8cm]{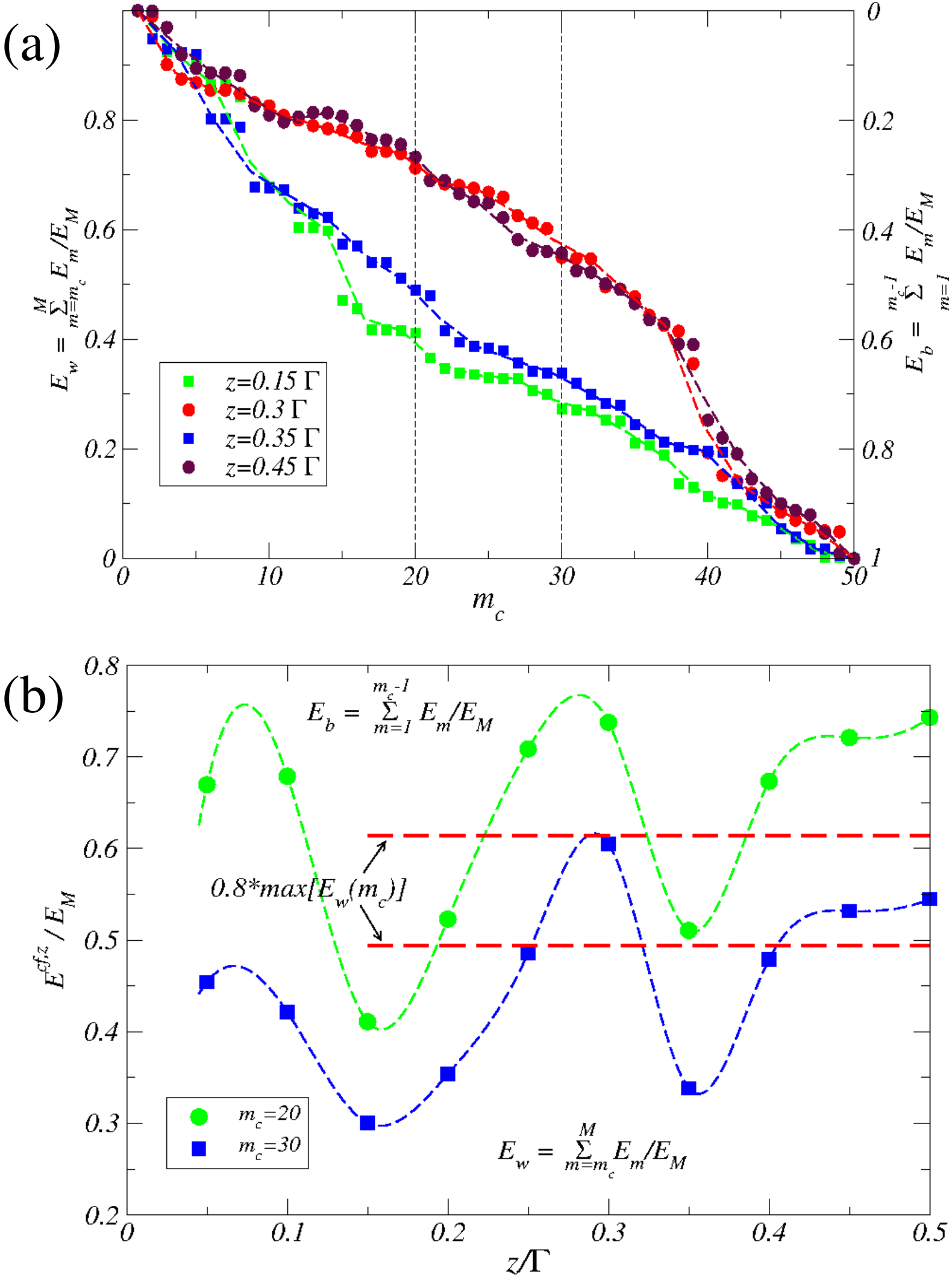}
\end{center} 
\caption{(Color online) {\bf Characterization of axial cross-flow energy based
on mode separation.} (a) Variation of the axial cross-flow energy component 
$E^{cf,z}_m(z) = E^{cf,z}_b(z)+E^{cf,z}_w(z)$ versus the cutoff wavenumber $m_c$
for two axial positions in the burst region ($z=0.3\Gamma$ and $0.45\Gamma$, 
circles) and in the background region ($z=0.15\Gamma$ and $0.35\Gamma$, squares). 
(b) Spatial variations of $E^{cf,z}_b(z) = \sum_{m=1}^{m_c-1} E^{cf,z}_m(z)$ and
$E^{cf,z}_w(z) = \sum_{m=m_c}^M E^{cf,z}_m(z)$ with $z$ for $m_c = 20$ (circles) 
and $30$ (squares). Regions above (below) each curve indicate $E_b$ ($E_w$).
The two horizontal dashed lines give the 80$\%$ threshold of the maximum values
of $[E_w(m_c=20)]$ (lower) and $[E_w(m_c=30)]$ (upper), respectively. In the
calculations the zero mode contribution is excluded. Due to normalization with
$E_M$ the sum must be unity.}
\label{fig:Ecf-mc}
\end{figure}

To better characterize the ring-burst patterns in relation to the background
wavy-like and general burst patterns, we devise a method based on the idea
of mode separation. Since a burst pattern includes modes of higher azimuthal 
wave numbers, we can decompose the axial component of the cross-flow energy
$E_m^{cf,z}(z,t)$ into two distinct subcomponents:
\begin{equation} \label{EQ:separate_sum_crossflow_m}
E^{cf,z}_w(z)+E^{cf,z}_b(z) 
= \sum_{m=1}^{m_c-1} E^{cf,z}_m(z) + \sum_{m=m_c}^M E^{cf,z}_m(z),
\end{equation}
where $E^{cf,z}_w(z)$ and $E^{cf,z}_b(z)$ denote the axial components
of the cross-flow energy associated with the background wavy-like and 
burst patterns, respectively, and $m_c$ is some cutoff mode number.
Note that the axisymmetric component of the cross-flow energy is excluded 
because it is significantly larger than all other components 
[Figs.~\ref{fig:Ecf-z}(a,b)]. We choose the normalization factor to be
the total cross-flow energy for all modes except $m=0$:
\begin{equation} \label{EQ:sum_crossflow_M}
E^{cf,z}_M = \sum_{m=1}^M E^{cf,z}_m.
\end{equation}
Figure~\ref{fig:Ecf-mc}(a) shows the basic $E_w$ and the burst contribution 
$E_b$ at different axial positions versus the cut-off wavenumber $m_c$.
We observe that the curves for $z$ positions within the burst region (circles,
e.g, at $z=0.3\Gamma \ \mbox{and} 0.45\Gamma$, 
Figs.~\ref{fig:contours_vort_full_without_zero} and \ref{fig:Ecf-z})
are higher than those in the wavy-like background (squares, e.g., 
at $z=0.15\Gamma \ \mbox{and} \ 0.35\Gamma$), where the former exhibit  
a rapid decrease in the energy to collapse with the latter for $m_c$ 
about $39$. The variations of $E_b$ and $E_w$ along the annulus length 
for two different cutoff wavenumbers [$m_c = 20,30$, as indicated by the 
vertical lines in Fig.~\ref{fig:Ecf-mc}(a)] are shown in Fig.~\ref{fig:Ecf-mc}(b).
Neglecting the differences in their magnitudes, the variations show  
qualitatively similar behaviors. Using a cutoff level with 80$\%$
of the maximum of the background contribution $[E_w(m_c)]$, we find a 
good agreement with the visible energy thresholds between the background and 
burst regions (cf. Fig.~\ref{fig:contours_vort_full_without_zero}). 
We thus see that, through some proper choice of the cutoff mode number,
the burst and non-burst regions can be distinguished by examining the 
axial cross-flow energy variations associated with the cutoff mode.

\subsection{Axial spacing and localization} \label{SUBSEC:axial_space}

\begin{figure}
\begin{center}
\includegraphics[width=0.9\linewidth,height=8cm]{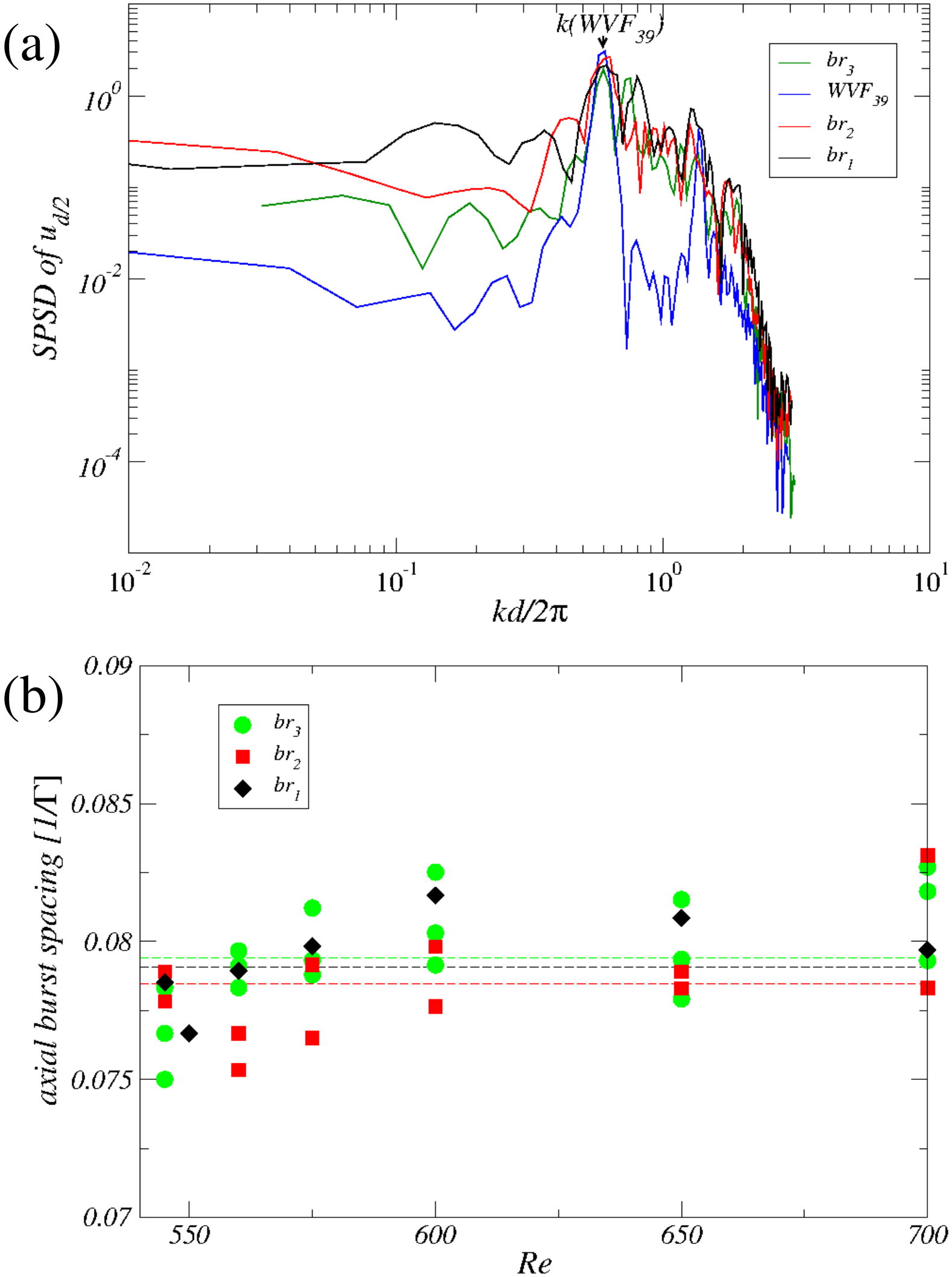}
\end{center}
\caption{(Color online) {\bf Axial spacing of ring bursts.}
(a) SPSD of the radial velocity $u_r$ at the midgap for $Re=545$, where $k$ 
is the axial wave number. (b) Axial spacing of the burst regions for order-$n$ 
ring bursts ($n\in\{1,2,3\}$), indicated by the horizontal lines. The average 
spacing is about 0.079, the axial spacing of two 
pairs of Taylor vortices constituting four vortex cells.}
\label{fig:SPSD_spacing}
\end{figure}

The axial ranges of distinct ring burst regions are approximately 
identical. Figure~\ref{fig:SPSD_spacing}(a) shows the spatial power 
spectral density (SPSD) of the radial velocity $u_r$ along a line at 
the midgap in the center of the bulk for the background flow and three
types of ring-burst patterns. We observe that the four curves coincide
at the first sharp peak determined by the wave number associated with 
the background flow WVF$_{39}$ that consists of 24 vortex pairs in the
axial direction. The corresponding axial wavelength and wave number are
$\lambda \approx 1.667$ and $k \approx 3.770$, respectively. 
In addition, several broadband peaks at higher wave numbers exist in 
the SPSD of the ring burst patterns, corresponding to a number of 
short wavelength bursts within the respective patterns. The broadband nature
at higher wave numbers are indicative of dominance of small-scale 
burst patterns. Figure~\ref{fig:SPSD_spacing}(b) shows the axial 
spacing associated with various ring burst patterns as a function of $Re$,
where each horizontal dashed line represents the averaged axial spacing 
for a particular ring burst pattern. The axial spacing is apparently 
independent of the value of $Re$ and of the order $n$ of the 
ring-burst pattern. The typical value of axial expansion agrees well
with the axial dimension of two-pair Taylor vortices that constitute 
four single vortex cells, which holds for all ring-burst flows that 
we have succeeded in uncovering. Analogous to the behavior of the 
cross-flow energy, this behavior is indicative of that of turbulent 
WVFs~\cite{TsFe2004,OMMSK1998,Don08a,Don2007}. We note that the size of 
only one pair of Taylor vortices (two cells) is too small to 
account for the observed range of axial expansion. This 
is consistent with the formation process of ring bursts. In particular,
any localized turbulent patch, after its generation, first expands in 
the axial direction (to four cells) before growing in the azimuthal 
direction. Whenever the burst region has expanded to a larger size in 
the axial direction, the closed ring structure is destroyed, leading 
to VSWBs.

\section{Conclusions and discussions} \label{sec:conclusions}

This paper provides a comprehensive numerical study of the quasi 
two-dimensional Taylor-Couette systems of radius ratio close but not
equal to unity, a regime that has not been studied previously. The 
relevant control or bifurcation parameter is the Reynolds number $Re$.
For small $Re$ values, TVF initially arises near one of the lids
in a single cell, extends and finally fills the bulk interior
completely. As $Re$ is increased the TVF loses its stability and
WVF emerges through a supercritical Hopf bifurcation. WVFs with 
high azimuthal wavenumbers, e.g., $m=39$, constitute a persistent
background flow, on top of which more complex flow structures develop, 
such as VSWBs that has been experimentally observed~\cite{CSA07}. 

The main result of this paper is the discovery of a new type of  
transient, intermittent state {\it en route} to turbulence
with increasing $Re$: ring bursts. They emerge when localized 
turbulent patches grow and close azimuthally, signifying a higher 
order instability. The ring bursts occur in various orders and 
the azimuthally closed burst regions possess axial expansion 
surrounded by the WVF background flow or more complex flows at 
high $Re$ values. The ring burst patterns differ characteristically
from the localized VSWB turbulent patches. The axial expansion of 
ring burst patterns of different orders correlates well with 
the size of the double-pair Taylor vortex structure, providing
plausible reason that the patterns are similar to the turbulent WVF 
structure~\cite{TsFe2004,OMMSK1998,Don08a,Don2007} that usually 
occurs through the whole annulus at higher Reynolds numbers.
We develop a mode separation method 
based on decomposing the cross-flow energy to distinguish between 
burst and non-burst patterns, where the former and latter are 
associated with higher and lower azimuthal modes, respectively.
We also find that the radial cross-flow energy changes significantly
in the presence of ring burst patterns. 
By examining the maximum value of the cross-flow
energy, we determine the onset of the ring-burst patterns at  
the critical Reynolds number of $Re_c \approx 537$, which is larger
than that for the onset of VSWBs~\cite{CSA07} (about 482). For 
ring burst patterns of different orders, their expansions in the
axial direction are nearly identical, and they tend to shift
along the axial direction. 

There are a number of differences between turbulence in the wide-gap
(e.g., radius ratio 0.5 to 0.8) and narrow-gap (e.g., radius ratio 
0.99) Taylor-Couette systems. Firstly, in the wide-gap case, 
the intensity distributions of turbulent fluctuations
are often uneven in the radial direction~\cite{Don2007,BrEc2013}, 
where more energetic turbulent fluctuations occur towards the inner 
cylinder wall. The regions near the inner cylinder thus exhibits 
stronger turbulence than in the region near the outer cylinder. 
This radial dependence of turbulent fluctuations is lost in the 
quasi two-dimensional, narrow-gap Taylor-Couette system, where  
turbulence is observed through the bulk in the radial direction.
Secondly, in wide-gap systems, phenomena such as turbulent 
streaks~\cite{Don2007}, small-scale G\"ortler vortices and 
herringbone-like streaks~\cite{Goe1954} can occur near both inner
and outer cylindrical walls. Examining the typical size of the small
G\"ortler vortices~\cite{Goe1954} reveals that, in the narrow-gap case these 
vortices have expansion larger than the radial width, excluding 
the possibility of generating turbulent streaks from small-scale
G\"ortler vortices. Indeed, our simulations do not reveal any kind 
of such small scale vortices. This might explain the loss of radial 
dependence of turbulence, as can be seen, e.g., from the cross-flow 
energy behavior in Fig.~\ref{fig:crossflow_all}. The mechanism that generates 
turbulence in quasi two-dimensional Taylor-Couette system is thus 
quite different from that in the three-dimensional system. In fact,
in narrow-gap systems turbulence emerges at the boundary layer of the
neighboring vortex cells almost immediately at any radial position.

Turbulence in quasi two-dimensional Taylor-Couette systems has 
different features than those in the planar systems as well. In particular,
firstly, in planar Couette flows there is a significant difference between
the velocities at the two walls, and asymmetry in the intensity
distribution of turbulent fluctuations is caused by the curvature 
effect. In the narrow-gap case, this difference is minimal.
Secondly, while there are ring burst patterns of turbulent 
bands in planar Couette flows~\cite{BaTu2005,TuBa2011}, the 
background flow is laminar, versus wavy-like patterns in quasi 
two-dimensional systems, where the former defines a threshold 
between basic state and turbulence but the latter is a threshold 
between an already complex state and turbulence. Except this
difference there is in fact a remarkable similarity between 
turbulence in both types of systems. For example, detailed
investigations~\cite{BaTu2005} of the laminar-turbulent boundary 
layer in planar Couette flows revealed isolated band states of 
turbulence in confined domains close to the global stability threshold.
Although these bands appear at various different angles, they are all 
parallel, which are remarkably analogous to our ring-burst patterns 
on an unrolled cylindrical surface (e.g., comparing 
Fig.~\ref{fig:contours_vort_full_without_zero} with Figs.~1 and $11$
in Ref.~\cite{TuBa2011}). In addition, the routes to turbulence are
similar: in both cases the turbulent bands grow out of a small 
localized turbulent spot that subsequently expands in some direction.
 
It may be challenging to detect ring-burst patterns experimentally 
as they coexist with other complex states such as VSWBs with similar
turbulent characteristics. Nonetheless, given that the Taylor-Couette 
system is a paradigm enabling well controlled experiments on complex 
vortex dynamics and turbulence, we hope our finding will 
stimulate further research of turbulence in quasi two-dimensional
regime of the system, a regime that has received little attention in 
spite of a large body of literature on Taylor-Couette flows in general.

\section*{Acknowledgments}

Y. D. was supported by Basic Science Research Program through the 
National Research Foundation of Korea (NRF) funded by the Ministry 
of Education, Science and Technology (NRF-2013R1A1A2010067). 
Y.-C. L. was supported by AFOSR under Grant No.~FA9550-12-1-0095.

\end{document}